%% file: Duplij-Fu-Guo_CryptoPolyRing-arxiv1.tex
\documentclass[aps,prd,onecolumn,12pt,longbibliography]{revtex4-2}
\usepackage{amsmath,amsthm}
\usepackage{amsfonts}
\usepackage{amssymb}
\usepackage{amsbsy,bm}

\theoremstyle{plain}
\newtheorem{theorem}{Theorem}

\theoremstyle{definition}

\theoremstyle{remark}

\newtheorem{example}[theorem]{Example}

\numberwithin{equation}{section}

\renewcommand{\mathit}{\bm}
\usepackage{natbib}

\begin{document}

\title{Cryptographic transformations over polyadic rings}

\author{\textbf{Steven Duplij}}
\email{douplii@uni-muenster.de, duplij@gmx.de, http://www.uni-muenster.de/IT.StepanDouplii}

\affiliation{
College of Information and Communication Engineering,
Harbin Engineering University,
150001 Harbin, China}

\affiliation{
Yantai Research Institute,
Harbin Engineering University,
265615 Yantai, China}

\affiliation{Center for Information Technology,
University of M\"unster,
48149 M\"unster,
Germany}

\author{\textbf{Qiang Guo}}
\email{guoqiang292004@163.com, guoqiang@hrbeu.edu.cn}

\affiliation{
College of Information and Communication Engineering,
Harbin Engineering University,
150001 Harbin, China}

\author{\textbf{Na Fu}}
\email{nafu@hrbeu.edu.cn}

\affiliation{
College of Information and Communication Engineering,
Harbin Engineering University,
150001 Harbin, China}

\date{December 14, 2025}

\begin{abstract}
\input{Duplij-Fu-Guo_CryptoPolyRing-abs}

\end{abstract}

\maketitle
\begin{small}
\tableofcontents
\end{small}%
\newpage
\input{Duplij-Fu-Guo_CryptoPolyRing-sw}
\newpage

\input{Duplij-Fu-Guo_CryptoPolyRing-arxiv.bbl}
\end{document}

%% file: Duplij-Fu-Guo_CryptoPolyRing-abs.tex

\noindent This article introduces a novel cryptographic paradigm based on
nonderived polyadic algebraic structures. Traditional cryptosystems rely on
binary operations within groups, rings, or fields, whose well-understood
properties can be exploited in cryptanalysis. To overcome these
vulnerabilities, we propose a shift to polyadic rings, which generalize
classical rings by allowing closed, nonderived operations of higher arity: an
$m$-ary addition and an $n$-ary multiplication.

The foundation of our approach is the construction of polyadic
integers---congruence classes of ordinary integers endowed with such $m$-ary
and $n$-ary operations. A key innovation is the parameter-to-arity mapping
$\Phi(a,b)=(m,n)$, which links the parameters $(a,b)$ defining a congruence
class to the specific arities required for algebraic closure. This mapping is
mathematically intricate: it is non-injective, non-surjective, and often
multivalued, meaning a single $(a,b)$ can correspond to multiple arity pairs
$(m,n)$, and vice versa. This complex, non-unique relationship forms the core
of the proposed cryptosystem's security.

We present two concrete encryption procedures that leverage this structure by
encoding plaintext within the parameters of polyadic rings and transmitting
information via polyadically quantized analog signals. In one method,
plaintext is linked to the additive arity $m_{i}$ and secured using the
summation of such signals; in the other, it is linked to a ring parameter
$a_{i}$ and secured using their multiplication. In both cases, the
\textquotedblleft quantized\textquotedblright\ nature of polyadic
operations---where only specific numbers of elements can be
combined---generates systems of equations that are straightforward for a
legitimate recipient with the correct key (knowledge of the specific polyadic
powers and functional dependencies used during encoding) but exceptionally
difficult for an attacker without it.

The resulting framework promises a substantial increase in cryptographic
security. The complexity of the $\Phi$-mapping, combined with the flexibility
in choosing polyadic powers and representative functions, creates a vast and
intricate key space. This makes brute-force attacks computationally infeasible
and complicates algebraic cryptanalysis, as the underlying nonderived polyadic
structures defy the linear properties and homomorphisms exploitable in
conventional binary algebraic systems. This work establishes the theoretical
foundation for this new class of encryption schemes, demonstrates their
feasibility through detailed examples, and highlights their potential for
constructing robust, next-generation cryptographic protocols.

%% file: Duplij-Fu-Guo_CryptoPolyRing-sw.tex

\section{\textbf{Introduction}}

Traditional algebraic structures, such as groups, rings, and fields, have long
served as the foundation for numerous encryption schemes
\cite{bertaccini,oriyano}. These schemes typically rely on operations defined
over sets with binary (two-place) addition and multiplication \cite{kat-lin}.
However, the inherent constraints and well-studied properties of these binary
structures can also make them vulnerable to cryptanalysis.

In this article, we explore a novel cryptographic paradigm grounded in
polyadic (multiary) algebraic structures \cite{duplij2022}. A polyadic ring
generalizes the classical notion of a ring by allowing operations of higher
arity: an $m$-ary addition and an $n$-ary multiplication \cite{lee/but}.
Crucially, these operations are nonderived, meaning they cannot be expressed
as iterations of binary operations. This shift to higher-arity algebras
introduces profound and complex structural properties, which can be harnessed
to create highly secure encryption mechanisms \cite{dup/guo}.

Our approach leverages the concept of polyadic integers \cite{dup2017a}, which
are constructed from congruence classes of ordinary integers but are endowed
with closed $m$-ary addition and $n$-ary multiplication. A key feature of this
construction is the parameter-to-arity mapping $\Phi(a,b)=(m,n)$, which links
the parameters $a$ and $b$ defining a congruence class to the specific arities
required for closure \cite{dup2019}. This mapping is mathematically intricate:
it is non-injective, non-surjective, and often multivalued, meaning a single
pair $(a,b)$ can correspond to multiple valid arity pairs $(m,n)$, and vice
versa \cite{duplij2022}.

This intricate, non-unique relationship forms the core of our proposed
cryptosystem \cite{dup/guo}. By encoding plaintext within the parameters of
these polyadic rings and transmitting information via polyadically quantized
analog signals, we can embed cryptographic keys within the very arity
structure of the operations. The decryption process requires not only the
intercepted signal amplitudes but also knowledge of the specific polyadic
powers and the functional dependencies used during encoding, which are not transmitted.

We detail two principal encryption procedures: one based on the summation of
polyadic signals and another on their multiplication. Both methods exploit the
\textquotedblleft quantized\textquotedblright\ nature of admissible operands
in polyadic operations only specific numbers of elements can be combined to
create systems of equations that are easy to solve for a legitimate recipient
with the correct key but exceptionally difficult for an attacker without it.

The resulting framework promises a significant increase in cryptographic
security. The complexity of the parameter-to-arity mapping, combined with the
flexibility in choosing polyadic powers and representative functions, creates
a vast and intricate key space. This makes brute-force attacks computationally
infeasible and algebraic cryptanalysis exceedingly challenging, as the
underlying structure defies the straightforward properties of conventional
binary algebraic systems. This article lays the theoretical groundwork for
this new class of polyadic encryption schemes \cite{dup/guo}, demonstrating
their feasibility through concrete examples and highlighting their potential
for creating robust, next-generation cryptographic protocols.

\section{\textbf{Preliminaries}}

We here remind the polyadic notations and conventions which will be used
below, for more details see \cite{duplij2022}.

Let $R$ be a set with two totally associative operations: $m$-ary addition
$\nu_{m}:R^{\times m}\rightarrow R$ and $n$-ary multiplication $\mu
_{n}:R^{\times n}\rightarrow R$, such that $\mathsf{G}_{m}=\left\langle
R\mid\nu_{m};assoc\right\rangle $ is a commutative $m$-ary group (each element
has its polyadic inverse, querelement), and $\mathsf{S}_{n}=\left\langle
R\mid\mu_{n};assoc\right\rangle $ is an $n$-ary semigroup. A polyadic or
$\left(  m,n\right)  $-ring is $\mathsf{R}_{m,n}=\left\langle R\mid\nu_{m}%
,\mu_{n};assoc,distr\right\rangle $ with the additional property, polyadic
distributivity, when $n$-ary multiplication $\mu_{n}$ \textquotedblleft
distributes\textquotedblright\ over $m$-ary addition $\nu_{m}$ \cite{lee/but}.
A polyadic ring is nonderived, if $m$-ary addition $\nu_{m}$ or $n$-ary
multiplication $\mu_{n}$ (or both) cannot be composed from binary operations.

A repeated composition of $\ell$ operations is called its power and denoted by
$\ell\in\mathbb{N}$. If an operation is binary, then obviously that the
admissable number of composed elements can be any $\ell+1$. This is not the
case for multiplace operations. In a polyadic $\left(  m,n\right)  $-ring we
can add the admissable number $p\left(  \ell_{m},m\right)  =\ell_{m}\left(
m-1\right)  +1$ elements and multiply the admissable number $p\left(  \ell
_{n},n\right)  =\ell_{n}\left(  n-1\right)  +1$ elements only, that is some
kind of \textquotedblleft quantization\textquotedblright, and we define%
\begin{align}
\nu_{m}^{\left[  \ell_{m}\right]  }\left[  r_{1},\ldots r_{\ell_{m}\left(
m-1\right)  +1}\right]   &  =\nu_{m}\left[  r_{1},r_{2},\ldots,r_{m-1},\nu
_{m}\left[  r_{m},r_{m+1}\ldots\nu_{m}\left[  \ldots r_{\ell_{m}\left(
m-1\right)  +1}\right]  \right]  \right]  ,\label{nm}\\
\mu_{n}^{\left[  \ell_{n}\right]  }\left[  r_{1},\ldots r_{\ell_{n}\left(
n-1\right)  +1}\right]   &  =\mu_{n}\left[  r_{1},r_{2},\ldots,r_{n-1},\mu
_{n}\left[  r_{n},r_{n+1}\ldots\mu_{n}\left[  \ldots r_{\ell_{n}\left(
n-1\right)  +1}\right]  \right]  \right]  , \label{mn}%
\end{align}
where $\ell_{m}$ is the polyadic power of $m$-ary addition, and $\ell_{n}$ is
the polyadic power of $n$-ary multiplication. So in the binary case any number
$\geq2$ of composed elements is admissable, because $p\left(  \ell
_{m},2\right)  =\ell_{m}+1$ ($p\left(  \ell_{n},2\right)  =\ell_{n}+1$), and
$\ell_{m},\ell_{n}\geq1$.

The simplest binary ring is the set of integer numbers $\mathsf{R}%
_{2,2}=\mathbb{Z}_{2,2}=\mathbb{Z}$. The set of positive integers
$\mathbb{Z}_{+}$ is the derived ternary ring, since a product of two elements
is in $\mathbb{Z}_{+}$ again, while the set of negative integers
$\mathbb{Z}_{-}$ is a nonderived ternary ring, because a product of two
elements is out of $\mathbb{Z}_{-},$and so the binary product is not closed. A
similar construction of nonderived polyadic rings was proposed in
\cite{dup2017a,dup2019}, where polyadic integers were introduced, as
representatives of congruence classes.

Let us denote the congruence class of an integer $a$ modulo integer $b$ (both
fixed) by%
\begin{align}
\left[  \left[  a\right]  \right]  _{b}  &  =\left\{  r^{\left[  a,b\right]
}\left(  k\right)  \mid k\in\mathbb{Z},,\ b\in\mathbb{N\setminus}\left\{
1\right\}  ,\ a\in\mathbb{Z}_{+},0\leq a\leq b-1\right\}  ,\label{ab}\\
r^{\left[  a,b\right]  }\left(  k\right)   &  =a+b\cdot k, \label{rk}%
\end{align}
where $r^{\left[  a,b\right]  }\left(  k\right)  $ is a generic representative
element of the class $\left[  \left[  a\right]  \right]  _{b}$, and we
excluded the trivial class $\left[  \left[  0\right]  \right]  _{1}%
=\mathbb{Z}$ having ordinary integers as the representatives $r^{\left[
0,1\right]  }\left(  k\right)  =k\in\mathbb{Z}$. The internal construction of
nontrivial congruence classes was never considered before \cite{dup2017a},
because with $b\geq2$ the binary addition and binary product of
representatives (\ref{rk}) are not closed (as the binary product of negative
numbers in the example above), and so no binary algebraic structure can be
defined of the set of representatives $\left\{  r^{\left[  a,b\right]
}\left(  k\right)  \right\}  $. Nevertheless, special operations of higher
arity for representatives can be closed also for nontrivial congruence
classes, which shows that a conguence class $\left[  \left[  a\right]
\right]  _{b}$ (\ref{ab}) is not only a set, but indeed a polyadic algebraic
structure, that is a polyadic ring \cite{dup2017a}.

To illustrate this, we first add $m$ representatives $r^{\left[  a,b\right]
}\left(  k\right)  $ and multiply $n$ ones (with fixed $a$, $b$) to obtain%
\begin{align}%
{\displaystyle\sum\limits_{i=1}^{m}}
\ r^{\left[  a,b\right]  }\left(  k_{i}\right)   &  =%
{\displaystyle\sum\limits_{i=1}^{m}}
\ \left(  a+b\cdot k_{i}\right)  =a\cdot m+b\cdot K\left(  m\right)
,\label{bk}\\%
{\displaystyle\prod\limits_{i=1}^{n}}
\ r^{\left[  a,b\right]  }\left(  k_{i}\right)   &  =%
{\displaystyle\prod\limits_{i=1}^{n}}
\ \left(  a+b\cdot k_{i}\right)  =a^{n}+b\cdot L\left(  n\right)  , \label{bl}%
\end{align}
where%
\begin{align}
K\left(  m\right)   &  =%
{\displaystyle\sum\limits_{i=1}^{m}}
\ k_{i},\label{km1}\\
L\left(  n\right)   &  =%
{\displaystyle\sum\limits_{i=1}^{n}}
\ a^{n-i}b^{i-1}s_{i}\left(  k_{1},\ldots,k_{n}\right)  , \label{ln1}%
\end{align}
and $s_{i}\left(  k_{1},\ldots,k_{n}\right)  $ are the symmetric polynomials
of $i$th degree with respect to $n$ variables
\begin{equation}
s_{i}\left(  k_{1},\ldots,k_{n}\right)  =\sum_{1\leq j_{1}<j_{2}\ldots
<j_{i-1}<j_{i}\leq n}k_{j_{1}},\ldots,k_{j_{i}} \label{sk}%
\end{equation}
For instance,%
\begin{align}
L\left(  3\right)   &  =a^{2}\left(  k_{1}+k_{2}+k_{3}\right)  +ab\left(
k_{1}k_{2}+k_{1}k_{3}+k_{2}k_{3}\right)  +b^{2}k_{1}k_{2}k_{3},\\
L\left(  4\right)   &  =a^{3}\left(  k_{1}+k_{2}+k_{3}+k_{4}\right)
+a^{2}b\left(  k_{1}k_{2}+k_{1}k_{3}+k_{1}k_{4}+k_{2}k_{3}+k_{2}k_{4}%
+k_{3}k_{4}\right) \nonumber\\
&  +ab^{2}\left(  k_{1}k_{2}k_{3}+k_{1}k_{2}k_{4}+k_{1}k_{3}k_{4}+k_{2}%
k_{3}k_{4}\right)  +b^{3}k_{1}k_{2}k_{3}k_{4},\\
&  \vdots\nonumber
\end{align}
and so on.

The main idea is possibility to close the addition (\ref{bk}) and
multiplication (\ref{bl}) for special values of $m$ and $n$ (till now they are
arbitrary). The closure of operation means that the result remains in the same
set, that is the r.h.s. of (\ref{bk}) and (\ref{bl}) should be representatives%
\begin{align}
a\cdot m+b\cdot K\left(  m\right)   &  =a+b\cdot k_{0},\label{amb}\\
a^{n}+b\cdot L\left(  n\right)   &  =a+b\cdot k_{00}, \label{anb}%
\end{align}
Because $k_{0},k_{00}$ should be integer (as for representatives (\ref{rk})),
the equations (\ref{amb}) and (\ref{anb}) become (after division by $b$)%
\begin{align}
k_{0}  &  =K\left(  m\right)  +I^{\left(  m\right)  }\left(  a,b\right)  ,\\
k_{00}  &  =L\left(  n\right)  +J^{\left(  n\right)  }\left(  a,b\right)  ,
\end{align}
where the introduces functions should be positive integer%
\begin{align}
I^{\left(  m\right)  }\left(  a,b\right)   &  =\frac{a\cdot m-a}{b}%
\in\mathbb{Z}_{+},\label{i}\\
J^{\left(  n\right)  }\left(  a,b\right)   &  =\frac{a^{n}-a}{b}\in
\mathbb{Z}_{+}. \label{j}%
\end{align}
This fixes the numbers of summands and multipliers%
\begin{align}
m  &  =m\left(  a,b\right)  ,\label{mm}\\
n  &  =n\left(  a,b\right)  , \label{nn}%
\end{align}
in the special way and allows us to close the addition (\ref{bk}) and
multiplication (\ref{bl}).

Now we take the congruence class $\left[  \left[  a\right]  \right]  _{b}$
(\ref{ab}) and define two new operations on its representatives: $m$-ary
addition $\nu_{m}$ and $n$-ary multiplication $\mu_{n}$ by%
\begin{align}
\nu_{m}\left[  r^{\left[  a,b\right]  }\left(  k_{1}\right)  ,r^{\left[
a,b\right]  }\left(  k_{2}\right)  ,\ldots,r^{\left[  a,b\right]  }\left(
k_{m}\right)  \right]   &  =r^{\left[  a,b\right]  }\left(  k_{1}\right)
+r^{\left[  a,b\right]  }\left(  k_{2}\right)  +\ldots+r^{\left[  a,b\right]
}\left(  k_{m}\right)  ,\label{n}\\
\mu_{n}\left[  r^{\left[  a,b\right]  }\left(  k_{1}\right)  ,r^{\left[
a,b\right]  }\left(  k_{2}\right)  ,\ldots,r^{\left[  a,b\right]  }\left(
k_{n}\right)  \right]   &  =r^{\left[  a,b\right]  }\left(  k_{1}\right)
\cdot r^{\left[  a,b\right]  }\left(  k_{2}\right)  \cdot\ldots\cdot
r^{\left[  a,b\right]  }\left(  k_{n}\right)  , \label{m}%
\end{align}
where the operations in the r.h.s., the binary addition $\left(  +\right)  $
and the binary multiplication $\left(  \cdot\right)  $, are in $\mathbb{Z}$.
Because both operations (\ref{n}) and (\ref{m}) are closed (only for special
arities (\ref{mm})--(\ref{nn})), we endow the set of representatives by
$\nu_{m}$ and $\mu_{n}$ to get a new algebraic structure, that is the polyadic
or $\left(  m,n\right)  $-ring (polyadic integers)%
\begin{equation}
\mathbb{Z}_{m,n}^{\left(  a,b\right)  }=\left\langle \left[  \left[  a\right]
\right]  _{b}\mid\nu_{m},\mu_{n};distr\right\rangle , \label{zab}%
\end{equation}
where polyadic distributivity follows from the binary distributivity, since
internal operations are in $\mathbb{Z}$. The dependences (\ref{mm}%
)--(\ref{nn}) can be written, as a special unified $2\longrightarrow2$ mapping%
\begin{equation}
\Phi\left(  a,b\right)  =\left(  m,n\right)  , \label{f}%
\end{equation}
which we call the parameter-to-arity (or arity shape) mapping, as the solution
to the consistency conditions (\ref{i})--(\ref{j}), and the functions
$I^{\left(  m\right)  }\left(  a,b\right)  $ and $J^{\left(  n\right)
}\left(  a,b\right)  $ are called arity shape invariants which fully
characterize the polyadic $\left(  m,n\right)  $-ring (\ref{zab}).

The table of the lowest values for the mapping $\Phi\left(  a,b\right)  $
(\ref{f}) is presented in \cite{dup2017a,dup/guo, duplij2022}. The parameteric
form of the integer solutions for (\ref{i})--(\ref{j}) is%
\begin{align}
m  &  =m_{u}\left(  a,b\right)  =1+u\cdot g\\
n  &  =n_{v}\left(  a,b\right)  =1+v\cdot\mathrm{ord}_{\ g}\left(  a\cdot
g\right)  ,\ \ \ u,v,g\in\mathbb{N}\mathbf{,}\\
g  &  =\frac{b}{\gcd\left(  a,b\right)  },
\end{align}
where $\mathrm{ord}_{y}x$ is the multiplicative order of $x$ modulo $y$, the
smallest positive integer $p$ such that $a^{p}\equiv1\operatorname{mod}y$, and
$\gcd\left(  a,b\right)  $ is the greatest common divisor for $a$ and $b$.

The central in the presented approach mapping $\Phi\left(  a,b\right)  $
(\ref{f}) is non-injective and non-surjective, also it is multivalued and not
unique. For instance, the following pairs $\left(  a,b\right)  $ (equivalence
classes) give multivalued arities $\left(  m,n\right)  $, we list some of them
only%
\begin{align}
\left(  196,245\right)   &  \mapsto\left(  51,21\right)  ,\ \left(
6,20\right)  ,\ \left(  51,15\right)  ,\ \left(  26,8\right)  ,\ \left(
46,23\right)  ,\\
\left(  610,705\right)   &  \mapsto\left(  988,69\right)  ,\ \left(
142,71\right)  ,\ \left(  1129,75\right)  ,\ \left(  142,81\right)  ,\ \left(
847,95\right)  .
\end{align}

Oppositely, the following multiple parameter pairs $\left(  a,b\right)  $
correspond to the same arity pair $\left(  m,n\right)  $%
\begin{align}
\left(  5,6\right)  ,\ \left(  9,18\right)  ,\ \left(  11,22\right)   &
\mapsto\left(  7,3\right)  ,\\
\left(  495,505\right)  ,\ \left(  504,707\right)  ,\ \left(  10,505\right)
&  \mapsto\left(  607,69\right)  .
\end{align}

It can be directly checked that in the above examples the arity shape
invariants $I^{\left(  m\right)  }\left(  a,b\right)  $ and $J^{\left(
n\right)  }\left(  a,b\right)  $ (\ref{i})--(\ref{j}) are interger, as they
should be. There is the set of pairs $\left(  a,b\right)  $ for which no
solutions for arity pairs $\left(  m,n\right)  $ exist at all, for example%
\begin{equation}
\left(  a,b\right)  =\left(  4,8\right)  ,\ \left(  10,16\right)  ,\ \left(
18,28\right)  ,\ \left(  12,24\right)  .
\end{equation}

Thus, the mapping $\Phi\left(  a,b\right)  $ (\ref{f}) is actually
multivalued, also non-injective and non-surjective.

This allows us to construct such reliable encryption schemes which are almost
impossible to decrypt or hack without special additional knowledge.

\section{\textbf{General polyadic encryption procedure}}

Encryption is the process of transforming plaintext into ciphertext to protect
its confidentiality. Only a holder of the correct cryptographic key can
reverse this process through decryption. Decryption is greatly simplified by
the fact that the texts usually have a similar algebraic nature and, at times,
originate from the same set (see, e.g. \cite{kat-lin,bertaccini,oriyano}).

We begin by representing the initial plaintext $T_{plain}$ (of the length $N$)
as a sequence of ordinary integers $\mathbb{Z}$, which can be always achieved
through a chosen encoding scheme, such that%
\begin{equation}
T_{plain}=\left\{  t_{1},t_{2},\ldots,t_{N}\right\}  ,\ \ \ \ t_{i}%
\in\mathbb{Z},\ \ i=1,\ldots,N, \label{t}%
\end{equation}
where, in our notation, $\mathbb{Z}=\mathcal{R}_{2,2}$ is a (binary) or
$\left(  2,2\right)  $-ring. The generalization to the polyadic case involves
replacing the binary ring $\mathbb{Z}$ with a polyadic ring $\mathcal{R}%
_{m,n}=\left\langle R\mid\nu_{m},\mu_{n}\right\rangle $, whose elements also
form the closed set $R$ with respect to $\nu_{m},\mu_{n}$. The information is
transfered to the receiver using signal series which have the parameters
connected with the polyadic rings $\mathcal{R}_{m_{i},n_{i}}=\left\langle
R\mid\nu_{m_{i}},\mu_{n_{i}}\right\rangle $ with the same underlying set $R$.
If the polyadic rings $\mathcal{R}_{m_{i},n_{i}}$ are polyadic integers
$\mathbb{Z}_{m_{i},n_{i}}^{\left(  a_{i},b_{i}\right)  }$, such technique can
be called polyadic discretization by analogy with the binary discretization in
which the parameters are ordinary integers $\mathbb{Z}$
\cite{oppenheim,ped/tro/per,berber,pro-man}. Here we consider the signal
amplitudes connected with the polyadic ring parameters $A_{i}\left(
a_{i},b_{i}\right)  \in\mathbb{Z}_{m_{i},n_{i}}^{\left(  a_{i},b_{i}\right)
}$, and such signals are called the continuous-time discrete-valued or
polyadic quantized analog signal, by analogy with the binary case
\cite{kar/ash/nij,mus/lis}.

This introduces the parameters of the polyadic ring $\mathbb{Z}_{m_{i},n_{i}%
}^{\left(  a_{i},b_{i}\right)  }$ as new variables. We thus propose a
cryptosystem where the extended plaintext is encrypted using the dyad (ordered
pair) of arities $\tbinom{m}{n}$, or the mixed dyad $\tbinom{a}{m}$, for each
entry in (\ref{t}) as a key component%
\begin{equation}
\mathbf{T}_{plain}=\left\{
\begin{array}
[c]{c}%
\tbinom{m_{1}}{n_{1}},\tbinom{m_{2}}{n_{2}},\ldots,\tbinom{m_{N}}{n_{N}}\\
\tbinom{a_{1}}{m_{1}},\tbinom{a_{2}}{m_{2}},\ldots,\tbinom{a_{N}}{m_{N}}%
\end{array}
\right.  \label{tp}%
\end{equation}

There are two possibilities to combine amplitudes: polyadic summation and
multiplication. The first case, which uses the dyads of arities $\tbinom
{m_{i}}{n_{i}}$, was briefly outlined in \cite{dup/guo}. In the second case we
cannot use multiplicative arities $n_{i}$, as parameters, because (as we will
see later on) the shape of equations depend on them. Therefore, we use another
polyadic ring parameter, that is $a_{i}$. Here we consider the both cases in
details and in unified way.

Let us encode the plaintext parameters (\ref{t}) by%
\begin{equation}
t_{i}\longrightarrow\left\{
\begin{array}
[c]{c}%
m_{i},\ \ \ \text{summation,}\\
a_{i},\ \ \ \text{multiplication,}%
\end{array}
\right.  ,\ \ \ \ \ i=1,\ldots,N, \label{ti}%
\end{equation}
but any functional dependence can be considered, which increases the security.
Then the ciphertext which is openly transferred to recepient using amplitudes
computed by sender will have the following form%
\begin{align}
\mathbf{T}_{cipher(sum)}  &  =\left\{  \tbinom{A_{sum,1}^{\left(  1\right)
},A_{sum,2}^{\left(  1\right)  },\ldots,A_{sum,M}^{\left(  1\right)  }}{n_{1}%
},\tbinom{A_{sum,1}^{\left(  2\right)  },A_{sum,2}^{\left(  2\right)  }%
,\ldots,A_{sum,M}^{\left(  2\right)  }}{n_{2}},\ldots,\tbinom{A_{sum,1}%
^{\left(  N\right)  },A_{sum,2}^{\left(  N\right)  },\ldots,A_{sum,M}^{\left(
N\right)  }}{n_{N}}\right\}  ,\label{tc1}\\
\mathbf{T}_{cipher(mult)}  &  =\left\{  \tbinom{A_{mult,1}^{\left(  1\right)
},A_{mult,2}^{\left(  1\right)  },\ldots,A_{mult,N}^{\left(  1\right)  }%
}{m_{1}},\tbinom{A_{mult,1}^{\left(  2\right)  },A_{mult,2}^{\left(  2\right)
},\ldots,A_{mult,N}^{\left(  2\right)  }}{m_{2}},\ldots,\tbinom{A_{mult,1}%
^{\left(  N\right)  },A_{mult,2}^{\left(  N\right)  },\ldots,A_{mult,N}%
^{\left(  N\right)  }}{m_{N}}\right\}  . \label{tc2}%
\end{align}

The recepient solves the system of equations for polyadically quantized
amplitudes $A_{sum,j}^{\left(  i\right)  }$ (or $A_{mult,j}^{\left(  i\right)
}$) and receives the initial parameters, and then the searched for plaintext
entries (\ref{t}). The additional row contain only arities which were not
taken into account for computations, and so they can be considered as a
polyadic analog for check bits: the recepient after computation can then
compute all other parameters using the parameter-arity shape correspondence
function $\Phi$ (\ref{f}) or Table 1 in \cite{dup/guo}. We can use one
polyadic check bit by adding all of them to prove the correctness of the decryption.

\section{\textbf{Polyadically quantized analog signals}}

Let us consider the normalized (in some sense) analog (continuous time) signal
$f^{\left(  \chi\right)  }\left(  t\right)  $ of the special species (or kind,
the natural $\chi\in\mathbb{N}$ numerates sine/cosine, triangular,
rectangular, etc.). Then the quantized analog signal
\cite{mus/lis,kar/ash/nij} is represented in the form%
\begin{equation}
\hat{X}_{j}^{\left(  \chi\right)  }\left(  t\right)  =A_{j}^{\left(
\chi\right)  }\cdot f^{\left(  \chi\right)  }\left(  t\right)  , \label{x}%
\end{equation}
where $\chi=1,2\ldots,P$ indexes the signal species with $P$ being the total
number of distinct signal species (or kinds), $j$ identifies an individual
signal within a given species $\chi$, $A_{j}^{\left(  \chi\right)  }$ is the
amplitude of that specific signal, and $f^{\left(  \chi\right)  }\left(
t\right)  $ is a characteristic waveform for species $\chi$. The standard
quantized amplitudes take value in the (binary) ring of ordinary integers
$A_{j}^{\left(  \chi\right)  }\in\mathbb{Z}$, and they were considered in
\cite{gho/kim/don,mus/lis}.

Here we introduce the polyadically quantized amplitudes \cite{dup/guo} which
belong to $\left(  m,n\right)  $-ring of polyadic integers $\mathbb{Z}_{m,n}$
\cite{dup2017a,duplij2022}. The presence of many additional parameters and
algebraic relations allows us to use various combination of polyadically
quantized amplitudes to transfer information with more higher level of
security than in the standard binary case, when amplitudes are quantized by
ordinary integers [encrypt binary].

Thus, if the amplitude is polyadically quantized, i.e. it belongs to the ring
of polyadic integers $A_{j}^{\left(  \chi\right)  }\in\mathbb{Z}_{m_{i},n_{i}%
}^{\left(  a_{i},b_{i}\right)  }$, we identify the $\chi$th species with the
fixed $i$th entry of the plaintext (\ref{t}), and the number of species $P$
with the length of the plaintext%
\begin{align}
\chi &  =i,\label{li}\\
P  &  =N. \label{ln}%
\end{align}
Therefore, the amplitude $A_{j}^{\left(  \chi=i\right)  }$ is in the set of
reprentatives of the coungruence class $\left[  \left[  a_{i}\right]  \right]
_{b_{i}}$, $b_{i}\in\mathbb{N}$, $0\leq a_{i}\leq b_{i}$, $i=1,\ldots,N$, such
that%
\begin{equation}
A_{j}^{\left(  \chi=i\right)  }=A_{j}^{\left(  \chi=i\right)  }\left(
a_{i},b_{i}\right)  =a_{i}+b_{i}\cdot k_{j}^{\left(  \chi=i\right)  }%
\in\mathbb{Z}_{m_{i},n_{i}}^{\left(  a_{i},b_{i}\right)  },\ \ \ \ k_{j}%
^{\left(  \chi=i\right)  }\in\mathbb{Z}, \label{ai}%
\end{equation}
where the arities of addition $m_{i}=m_{i}\left(  a_{i},b_{i}\right)  $ and
multiplication $n_{i}=n_{i}\left(  a_{i},b_{i}\right)  $ of the polyadic rings
$\mathbb{Z}_{m_{i},n_{i}}^{\left(  a_{i},b_{i}\right)  }$ can be computed
using (\ref{i})--(\ref{j}) or Table 1 of \cite{dup/guo}.

Recall, that in the $\left(  m_{i},n_{i}\right)  $-ring we can add only the
admissible number of ring elements equal to $\ell_{m_{i}}\left(
m_{i}-1\right)  +1$ and multiply only $\ell_{n_{i}}\left(  n_{i}-1\right)  +1$
elements, where $\ell_{m_{i}}$ and $\ell_{n_{i}}$ are polyadic powers of
addition and multiplication, correspondingly (see (\ref{nm})--(\ref{mn})).
Therefore, we are forced to distinguish the case of addition of signals and
the case of multiplication of signals of each fixed species $\chi=i$.

The main idea is to transfer by open channels so many polyadic powers for each
entry $t_{i}$ of the plaintext (\ref{t}), which will be sufficient to recover
the additive or multiplicative arities. At first glance, the number of
polyadic powers for each $i$ is two, just to transfer the pair $\left(
a_{i},b_{i}\right)  $, because $m_{i}=m_{i}\left(  a_{i},b_{i}\right)  $ (or
$n_{i}=n_{i}\left(  a_{i},b_{i}\right)  $). But the arity shape function
(\ref{f}) is has not very good behaviour for direct encodings: it is
non-injective and non-surjective simultaneously. So we need different polyadic
powers of amplitude $M=3$ in (\ref{tc1})--(\ref{tc2}) for each entry $i$ (see
(\ref{nm})), together with the polyadic check bit $n_{i}$ (or $m_{i}$) to
restore four parameters. Then, with given 3 parameters $\left(  a_{i}%
,b_{i}\right)  $ and $n_{i}$ (or $m_{i}$), the forth parameter, the check bit,
$m_{i}$ (or $n_{i}$) will be determined uniquely by the arity shape function
(\ref{f}) or Table 1 of \cite{dup2017a,dup/guo}. Moreover, there are possible
the mixed cases, when the column in (\ref{tc1})--(\ref{tc2}) can be
interchanged to be described by polyadic summation or multiplication
separately, following some additional rule. This cane increase security of the
signal transferring as well.

\section{\textbf{Encryption by summation of signals}}

Let us examine, which sums of polyadically quantized analog signals $\hat
{X}_{j}^{\left(  \chi\right)  }\left(  t\right)  $ (\ref{x}) of the same
species $\chi$ can be prepared. Because we can add only admissible number
$\ell_{m_{i}}\left(  m_{i}-1\right)  +1$ of amplitudes $A_{j}^{\left(
\chi=i\right)  }\in\mathbb{Z}_{m_{i},n_{i}}^{\left(  a_{i},b_{i}\right)  }$
(\ref{nm}), for signals we have, taking into account the identification of
species $\chi=i$ (\ref{li})%
\begin{equation}
\hat{X}_{sum}^{\left(  \chi=i\right)  }\left(  t\right)  =\nu_{m_{i}}^{\left[
\ell_{m_{i}}\right]  }\left[  A_{1}^{\left(  \chi=i\right)  }A_{2}^{\left(
\chi=i\right)  },\ldots,A_{\ell_{m_{i}}\left(  m_{i}-1\right)  +1}^{\left(
\chi=i\right)  }\right]  \cdot f^{\left(  \chi=i\right)  }\left(  t\right)
=A_{sum,\ell_{m_{i}}}^{\left(  \chi=i\right)  }\cdot f^{\left(  \chi=i\right)
}\left(  t\right)  ,
\end{equation}
where $\nu_{m_{i}}^{\left[  \ell_{m_{i}}\right]  }$ is the $\ell_{m_{i}}%
$-polyadic power of addition (\ref{nm}) in the polyadic ring $\mathbb{Z}%
_{m_{i},n_{i}}^{\left(  a_{i},b_{i}\right)  }$, and using (\ref{nm}) the total
summation amplitude becomes%
\begin{equation}
A_{sum,\ell_{m_{i}}}^{\left(  \chi=i\right)  }=\sum_{j=1}^{\ell_{m_{i}}\left(
m_{i}-1\right)  +1}A_{j}^{\left(  \chi=i\right)  }. \label{am}%
\end{equation}

Then we recall that the amplitudes $A_{j}^{\left(  \chi=i\right)  }$ are
polyadic integer numbers, that is in the polyadic ring $\mathbb{Z}%
_{m_{i},n_{i}}^{\left(  a_{i},b_{i}\right)  }$, and have the manifest form
(\ref{ai}) as representatives of the congriuence class $\left[  \left[
a_{i}\right]  \right]  _{i}$, and therefore the summation amplitudes
(\ref{am}) for each species $\chi=i$ can be presented as (cf.(\ref{amb}))%
\begin{equation}
A_{sum,\ell_{m_{i}}}^{\left(  \chi=i\right)  }=a_{i}\cdot\left(  \ell_{m_{i}%
}\left(  m_{i}-1\right)  +1\right)  +b_{i}\cdot K\left(  m_{i},\ell_{m_{i}%
}\right)  , \label{as}%
\end{equation}
where%
\begin{equation}
K\left(  m_{i},\ell_{m_{i}}\right)  =\sum_{j=1}^{\ell_{m_{i}}\left(
m_{i}-1\right)  +1}k_{j}^{\left(  \chi=i\right)  }, \label{k}%
\end{equation}
and $i$ is the number of the entry $t_{i}$ in the initial plaintext (\ref{t}),
$m_{i}$ is the arity of addition of polyadic numbers and $\ell_{m_{i}}$ is the
polyadic power of $m_{i}$-ary addition (\ref{nm}). Initially, the dependence
$k_{j}^{\left(  \chi=i\right)  }\left(  j\right)  $ is arbitrary.

As noted above, to securely transfer the additive arity $m_{i}$, each
plaintext entry $t_{i}$ must be encoded using three polyadically quantized
signals with amplitudes defined in (\ref{as}). This can be achieved by
selecting amplitudes with three distinct, arbitrary additive polyadic powers
$\ell_{m_{i}}^{\prime},\ell_{m_{i}}^{\prime\prime},\ell_{m_{i}}^{\prime
\prime\prime}$ (\ref{nm}), together with the multiplicative arity $n_{i}$
(second row in (\ref{tc1}), as the polyadic check bit, and by arbitrarily
establishing the specific functional dependence of the representative number
$k_{j}^{\left(  \chi=i\right)  }\left(  j\right)  $ in (\ref{k}).

Because all the parameters of each plaintext entry $t_{i}$ in (\ref{t}) can be
chosen separately, we consider only calculation of only one of them. If some
of parameters are taken the same for the whole plaintext, which can be agreed
before, then plaintexts can be distinguished by these parameters.

\begin{example}
Let us consider the plaintext consisting of four entries $\chi=i=1,\ldots,4$,
as
\begin{equation}
T_{plain(sent)}=\left\{  t_{1},t_{2},t_{3},t_{4}\right\}  =\left\{
m_{1},m_{2},m_{3},m_{4}\right\}  =\left\{  15,18,43,8\right\}  , \label{t1}%
\end{equation}
where $m_{i}$ are additive arities of the polyadic rings $\mathbb{Z}%
_{15,13}^{\left(  5,7\right)  }$, $\mathbb{Z}_{18,5}^{\left(  13,17\right)  }%
$, $\mathbb{Z}_{43,13}^{\left(  8,21\right)  }$ and $\mathbb{Z}_{8,4}^{\left(
2,7\right)  }$, correspondingly. So the dyad plaintext (\ref{tp}) becomes%
\begin{equation}
\mathbf{T}_{plain}=\left\{  \dbinom{15}{13},\dbinom{18}{5},\dbinom{43}%
{13},\dbinom{8}{4}\right\}  . \label{tp1}%
\end{equation}
For the whole plaintext we choice the same polyadic powers $\ell_{m_{i}%
}^{\prime}=\ell^{\prime}=2$, $\ell_{m_{i}}^{\prime\prime}=\ell^{\prime\prime
}=3$, $\ell_{m_{i}}^{\prime\prime\prime}=\ell^{\prime\prime\prime}=5$ and the
same functional dependence of the representative number $k_{j}^{\left(
\chi=i\right)  }$ in (\ref{k}) as%
\begin{equation}
k_{j}^{\left(  \chi=i\right)  }=3j^{2}+4j-5. \label{kj}%
\end{equation}

In this case the sum (\ref{k}) becomes%
\begin{equation}
K\left(  m_{i},\ell_{m_{i}}\right)  =\frac{1}{2}\left(  \ell_{m_{i}}\left(
m_{i}-1\right)  +1\right)  \left[  2\left(  \ell_{m_{i}}\left(  m_{i}%
-1\right)  +1\right)  ^{2}+7\left(  \ell_{m_{i}}\left(  m_{i}-1\right)
+1\right)  -5\right]  . \label{km}%
\end{equation}
Then, for $\ell^{\prime}=2,\ell^{\prime\prime}=3,\ell^{\prime\prime\prime}=5$
the summation amplitudes have the form%
\begin{align}
A_{sum,\ell_{m_{i}}=2}^{\left(  \chi=i\right)  }  &  =\left(  a_{i}+\frac
{1}{2}b_{i}\left(  14m_{i}+2\left(  2m_{i}-1\right)  ^{2}-12\right)  \right)
\left(  2m_{i}-1\right)  ,\label{a1}\\
A_{sum,\ell_{m_{i}}=3}^{\left(  \chi=i\right)  }  &  =\left(  a_{i}+\frac
{1}{2}b_{i}\left(  21m_{i}+2\left(  3m_{i}-2\right)  ^{2}-19\right)  \right)
\left(  3m_{i}-2\right)  ,\label{a2}\\
A_{sum,\ell_{m_{i}}=5}^{\left(  \chi=i\right)  }  &  =\left(  a_{i}+\frac
{1}{2}b_{i}\left(  35m_{i}+2\left(  5m_{i}-4\right)  ^{2}-33\right)  \right)
\left(  5m_{i}-4\right)  , \label{a3}%
\end{align}
where $\chi=i=1,2,3,4$, is the number of entry in the plaintext (\ref{t}).

It follows from (\ref{as}), (\ref{kj}) and (\ref{km}), that the summation
amplitudes numerically are%

\begin{align}
\mathbb{Z}_{15,13}^{\left(  5,7\right)  }  &  :A_{sum,\ell_{m_{i}}=2}^{\left(
\chi=i=1\right)  }=190965,\ A_{sum,\ell_{m_{i}}=3}^{\left(  \chi=i=1\right)
}=601312,\ A_{sum,\ell_{m_{i}}=5}^{\left(  \chi=i=1\right)  }=2627994,\\
\mathbb{Z}_{18,5}^{\left(  13,17\right)  }  &  :A_{sum,\ell_{m_{i}}%
=2}^{\left(  \chi=i=2\right)  }=800730,\ A_{sum,\ell_{m_{i}}=3}^{\left(
\chi=i=2\right)  }=2549690,\ A_{sum,\ell_{m_{i}}=5}^{\left(  \chi=i=2\right)
}=11250477,\\
\mathbb{Z}_{43,13}^{\left(  8,21\right)  }  &  :A_{sum,\ell_{m_{i}}%
=2}^{\left(  \chi=i=3\right)  }=13423880,\ A_{sum,\ell_{m_{i}}=3}^{\left(
\chi=i=3\right)  }=44195873,\ A_{sum,\ell_{m_{i}}=5}^{\left(  \chi=i=3\right)
}=200535455,\\
\mathbb{Z}_{8,4}^{\left(  2,7\right)  }  &  :A_{sum,\ell_{m_{i}}=2}^{\left(
\chi=i=3\right)  }=3493,\ A_{sum,\ell_{m_{i}}=3}^{\left(  \chi=i=3\right)
}=9295,\ A_{sum,\ell_{m_{i}}=5}^{\left(  \chi=i=3\right)  }=34696.
\end{align}

Thus, the ciphertext (\ref{tc1}) in the dyad form sent to the recepient by
open channel is%
\begin{align}
\mathbf{T}_{cipher(sum)}  &  =\left\{  \dbinom{190965,\ 601312,\ 2627994}%
{13},\ \dbinom{800730,\ 2549690,\ 11250477}{5}\right.  ,\nonumber\\
&  \left.  \dbinom{13423880,\ 44195873,\ 200535455}{13},\ \dbinom
{3493,\ 9295,\ 34696}{4}\right\}  . \label{tc}%
\end{align}
The recepient knows the system of equations for parameters (\ref{a1}%
)--(\ref{a3}) together with the polyadic powers $\ell=2,3,5$ (\ref{nm}), for
each plaintext entry $\chi=i=1,2,3,4$, he inserts the amplitude triples into
the system and solves it in integers for $\left(  a_{i},b_{i},m_{i}\right)  $,
to obtain the \textquotedblleft hacked\textquotedblright\ ciphertext as the
dyads $\tbinom{a_{i},b_{i},m_{i}}{n_{i}}$ in the following form%
\begin{equation}
\mathbf{T}_{ckacked(sum)}=\left\{  \dbinom{5,\ 7,\ 15}{13},\ \dbinom
{13,\ 17,\ 18}{5},\ \dbinom{8,\ 21,\ 43}{13},\ \dbinom{2,\ 7,\ 8}{4}\right\}
. \label{tcr}%
\end{equation}

Then, the recepient uses the check bits $n_{i}$ (sent him openly) to prove
that all the obtained and computed parameters actually satisfy the
parameter-arity shape mapping $\Phi\left(  a_{i},b_{i}\right)  =\left(
m_{i},n_{i}\right)  $ (\ref{f}), for smaller arities it is in the Table 1 of
\cite{dup2017a,dup/guo}, such that%
\begin{align}
\Phi\left(  5,7\right)   &  =\left(  8,7\right)  ,\ \ \ \frame{$\left(
15,13\right)  $},\ldots,\label{f1}\\
\Phi\left(  13,17\right)   &  =\frame{$\left(  18,5\right)  $},\ \ \ \left(
35,9\right)  ,\ldots,\\
\Phi\left(  8,21\right)   &  =\left(  22,7\right)  ,\ \ \ \frame{$\left(
43,13\right)  $},\ldots,\\
\Phi\left(  2,7\right)   &  =\frame{$\left(  8,4\right)  $},\ \ \ \left(
15,17\right)  ,\ldots, \label{f4}%
\end{align}
where the framed pairs $\left(  m_{i},n_{i}\right)  $ are chosen arities,
because $\Phi$ is multivalued, and the check bits $n_{i}$ indicate which pair
to use. For consistency of the mappings (\ref{f1})--(\ref{f4}), we present the
corresponding arity shape invariants $I^{\left(  m_{i}\right)  }\left(
a_{i},b_{i}\right)  ,J^{\left(  n_{i}\right)  }\left(  a_{i},b_{i}\right)  $ (
) which are all integer, as they should be%
\begin{align}
&  \left(  I^{\left(  8\right)  }\left(  5,7\right)  =5,J^{\left(  7\right)
}\left(  5,7\right)  =11160\right)  ,\ \ \left(  I^{\left(  15\right)
}\left(  5,7\right)  =10,J^{\left(  13\right)  }\left(  5,7\right)
=174386160\right)  ,\\
&  \left(  I^{\left(  18\right)  }\left(  13,17\right)  =13,J^{\left(
5\right)  }\left(  13,17\right)  =21840\right)  ,\ \ \left(  I^{\left(
35\right)  }\left(  13,17\right)  =26,J^{\left(  9\right)  }\left(
13,17\right)  =623794080\right)  ,\\
&  \left(  I^{\left(  22\right)  }\left(  8,21\right)  =16,J^{\left(
7\right)  }\left(  8,21\right)  =99864\right)  ,\ \ \left(  I^{\left(
43\right)  }\left(  8,21\right)  =8,J^{\left(  13\right)  }\left(
8,21\right)  =26178848280\right)  ,\\
&  \left(  I^{\left(  8\right)  }\left(  2,7\right)  =2,J^{\left(  4\right)
}\left(  2,7\right)  =2\right)  ,\ \ \left(  I^{\left(  15\right)  }\left(
2,7\right)  =4,J^{\left(  17\right)  }\left(  2,7\right)  =18\right)  .
\end{align}

Finally, this consistency check gives the initial plaintext (\ref{t1}) as the
set of the arities of addition $m_{i}$ in the corresponding four polyadic
rings $\mathbb{Z}_{m_{i},n_{i}}^{\left(  a_{i},b_{i}\right)  }=\mathbb{Z}%
_{15,13}^{\left(  5,7\right)  }$,$\mathbb{Z}_{18,5}^{\left(  13,17\right)  }%
$,$\mathbb{Z}_{43,13}^{\left(  8,21\right)  }$,$\mathbb{Z}_{8,4}^{\left(
2,7\right)  }$%
\begin{equation}
T_{plain(received)}=\left(  15,18,43,8\right)  =T_{plain(sent)}.
\end{equation}

\end{example}

\section{\textbf{Encryption by multiplication of signals}}

Here we consider the products of polyadically quantized analog signals
$\hat{X}_{j}^{\left(  \chi\right)  }\left(  t\right)  $ (\ref{x}) of the same
species $\chi=i$. We can multiply only admissible number $\ell_{n_{i}}\left(
n_{i}-1\right)  +1$ of amplitudes $A_{j}^{\left(  \chi=i\right)  }%
\in\mathbb{Z}_{m_{i},n_{i}}^{\left(  a_{i},b_{i}\right)  }$ (\ref{mn}), for
signals we have as follows%
\begin{equation}
\hat{X}_{mult}^{\left(  \chi=i\right)  }\left(  t\right)  =\mu_{m_{i}%
}^{\left[  \ell_{n_{i}}\right]  }\left[  A_{1}^{\left(  \chi=i\right)  }%
A_{2}^{\left(  \chi=i\right)  },\ldots,A_{\ell_{n_{i}}\left(  n_{i}-1\right)
+1}^{\left(  \chi=i\right)  }\right]  \cdot f^{\left(  \chi=i\right)  }\left(
t\right)  =A_{mult,\ell_{n_{i}}}^{\left(  \chi=i\right)  }\cdot f^{\left(
\chi=i\right)  }\left(  t\right)  ,
\end{equation}
where $\mu_{m_{i}}^{\left[  \ell_{m_{i}}\right]  }$ is the $\ell_{n_{i}}%
$-polyadic power of multiplication in the polyadic ring $\mathbb{Z}%
_{m_{i},n_{i}}^{\left(  a_{i},b_{i}\right)  }$, and using (\ref{mn}) the total
multiplication amplitude becomes%
\begin{equation}
A_{mult,\ell_{m_{i}}}^{\left(  \chi=i\right)  }=%
{\displaystyle\prod\limits_{j=1}^{\ell_{n_{i}}\left(  n_{i}-1\right)  +1}}
A_{j}^{\left(  \chi=i\right)  }. \label{an}%
\end{equation}

The amplitudes $A_{j}^{\left(  \chi=i\right)  }$ are polyadic interger numbers
from $\mathbb{Z}_{m_{i},n_{i}}^{\left(  a_{i},b_{i}\right)  }$, and have the
manifest form (\ref{ai}) as representatives of the congriuence class $\left[
\left[  a_{i}\right]  \right]  _{i}$. But now the multiplication amplitudes
(\ref{an}) for each species $\chi=i$ cannot be presented in the closed formula
as in the summation case (\ref{as}).Nevertheless, by analogy with (\ref{anb})
we obtain%
\begin{equation}
A_{mult,\ell_{m_{i}}}^{\left(  \chi=i\right)  }=a_{i}^{\ell_{n_{i}}\left(
n_{i}-1\right)  +1}+b_{i}\cdot L\left(  n_{i},\ell_{n_{i}}\right)  ,
\label{al}%
\end{equation}
where $L\left(  n_{i},\ell_{n_{i}}\right)  $ can be expressen through
elementary symmetric polynomials over $k_{j}^{\left(  \chi=i\right)  }$ in
each concrete case. Here $i$ is the number of the entry $t_{i}$ in the initial
plaintext (\ref{t}), $n_{i}$ is the arity of multiplication of polyadic
numbers and $\ell_{n_{i}}$ is the polyadic power of $n_{i}$-ary multiplication
(\ref{mn}). Thus, informally, we can write%
\begin{align}
L\left(  n_{i},\ell_{n_{i}}\right)   &  =a_{i}^{\ell_{n_{i}}\left(
n_{i}-1\right)  }S_{1}\left(  k\right)  +a_{i}^{\ell_{n_{i}}\left(
n_{i}-1\right)  -1}b_{i}\cdot S_{2}\left(  k\right)  +a_{i}^{\ell_{n_{i}%
}\left(  n_{i}-1\right)  -2}b_{i}^{2}\cdot S_{3}\left(  k\right) \nonumber\\
&  +a_{i}^{\ell_{n_{i}}\left(  n_{i}-1\right)  -3}b_{i}^{3}\cdot S_{4}\left(
k\right)  +\ldots+b_{i}^{\ell_{n_{i}}\left(  n_{i}-1\right)  }\cdot
S_{\ell_{n_{i}}\left(  n_{i}-1\right)  +1}\left(  k\right)  , \label{ls}%
\end{align}
where the elementary symmetric polynomials $S_{r}\left(  k\right)  $,
$r=1,\ldots,\ell_{n_{i}}\left(  n_{i}-1\right)  +1$, are ($k_{j}%
=k_{j}^{\left(  \chi=i\right)  }$)%
\begin{align}
S_{1}\left(  k\right)   &  =\sum_{j=1}^{\ell_{n_{i}}\left(  n_{i}-1\right)
+1}k_{j},\ \ \ \ell_{n_{i}}\left(  n_{i}-1\right)  +1\ \text{terms}%
,\label{s1}\\
S_{2}\left(  k\right)   &  =\sum_{1\leq j_{1}<j_{2}\leq\ell_{n_{i}}\left(
n_{i}-1\right)  +1}^{\ell_{n_{i}}\left(  n_{i}-1\right)  +1}k_{j_{1}}k_{j_{2}%
},\ \ \ \dbinom{\ell_{n_{i}}\left(  n_{i}-1\right)  +1}{2}\ \text{terms},\\
S_{3}\left(  k\right)   &  =\sum_{1\leq j_{1}<j_{2}<j_{3}\leq\ell_{n_{i}%
}\left(  n_{i}-1\right)  +1}^{\ell_{n_{i}}\left(  n_{i}-1\right)  +1}k_{j_{1}%
}k_{j_{2}}k_{j_{3}},\ \ \ \dbinom{\ell_{n_{i}}\left(  n_{i}-1\right)  +1}%
{3}\ \text{terms},\\
&  \vdots
\end{align}%
\begin{align}
S_{\ell_{n_{i}}\left(  n_{i}-1\right)  }\left(  k\right)   &  =\sum_{1\leq
j_{1}<j_{2}<\ldots<j_{\ell_{n_{i}}\left(  n_{i}-1\right)  }\leq\ell_{n_{i}%
}\left(  n_{i}-1\right)  +1}^{\ell_{n_{i}}\left(  n_{i}-1\right)  +1}k_{j_{1}%
}k_{j_{2}}\ldots k_{j_{\ell_{n_{i}}\left(  n_{i}-1\right)  }},\ \ \ \ell
_{n_{i}}\left(  n_{i}-1\right)  +1\ \text{terms},\\
S_{\ell_{n_{i}}\left(  n_{i}-1\right)  +1}\left(  k\right)   &  =k_{j_{1}%
}k_{j_{2}}\ldots k_{j_{\ell_{n_{i}}\left(  n_{i}-1\right)  +1}}%
,\ \ \ 1\ \text{term}. \label{sr}%
\end{align}

For instance, for ternary multiplication $n_{i}=3$ of second power
$\ell_{n_{i}}=2$, we obtain%
\begin{align}
&  L\left(  n_{i}=3,\ell_{n_{i}}=2\right)  =a_{i}^{4}\left(  k_{1}+k_{2}%
+k_{3}+k_{4}+k_{5}\right) \nonumber\\
&  +a_{i}^{3}b_{i}\left(  k_{1}k_{2}+k_{1}k_{3}+k_{1}k_{4}+k_{1}k_{5}%
+k_{2}k_{3}+k_{2}k_{4}+k_{2}k_{5}+k_{3}k_{4}+k_{3}k_{5}+k_{4}k_{5}\right)
\nonumber\\
&  +a_{i}^{2}b_{i}^{2}\left(  k_{1}k_{2}k_{3}+k_{1}k_{2}k_{4}+k_{1}k_{2}%
k_{5}+k_{1}k_{3}k_{4}+k_{1}k_{3}k_{5}+k_{1}k_{4}k_{5}\right. \nonumber\\
&  \left.  +k_{2}k_{3}k_{4}+k_{2}k_{3}k_{5}+k_{2}k_{4}k_{5}+k_{3}k_{4}%
k_{5}\right) \nonumber\\
&  +a_{i}b_{i}^{3}\left(  k_{1}k_{2}k_{3}k_{4}+k_{1}k_{2}k_{3}k_{5}+k_{1}%
k_{2}k_{4}k_{5}+k_{1}k_{3}k_{4}k_{5}+k_{2}k_{3}k_{4}k_{5}\right) \nonumber\\
&  +b_{i}^{4}k_{1}k_{2}k_{3}k_{4}k_{5} \label{l3}%
\end{align}

To enable the secure transfer of the multiplicative arity $n_{i}$, every
plaintext entry $t_{i}$ in (\ref{t}) is encoded into three polyadically
quantized signals. The amplitudes of these signals, which follow the form
prescribed in (\ref{al}), are set by making three arbitrary selections: first,
choosing two distinct multiplicative polyadic powers $\ell_{n_{i}}^{\prime
},\ell_{n_{i}}^{\prime\prime}$ (\ref{mn}), and the additive arity $m_{i}$ as a
check bit, and second, arbitrarily defining how the representative number
$k_{j}^{\left(  \chi=i\right)  }$ in (\ref{k}).

\begin{example}
Because the encryption by multiplication of amplitudes is more cumbersome than
by addition of them, we consider the initial plaintext consisting of three
entries. This would be sufficient to understand the general procedure and
difference from the previous (summation) case. We will take as multiplicative
polyadic powers $\ell_{n_{i}}^{\prime}=1$, $\ell_{n_{i}}^{\prime\prime}=2$
(\ref{mn}), and the simplest shape of the representative number function as%
\begin{equation}
k_{j}^{\left(  \chi=i\right)  }=j. \label{ki}%
\end{equation}

Since the product shape of amplitudes (\ref{an})--(\ref{ls}) depends on the
multiplicative arity $n_{i}$ (the number of terms in (\ref{ls}) is
$\ell_{n_{i}}\left(  n_{i}-1\right)  +1$ (\ref{mn})), $n_{i}$ cannot be the
seeked variable in equations. Therefore, in the multiplicative case we will
take variables $a_{i}$ of $\mathbb{Z}_{m_{i},n_{i}}^{\left(  a_{i}%
,b_{i}\right)  }$ as the plaintext parameter $t_{i}=a_{i}$. In this example we
take $n_{i}$ as its minimal value $n=3$, and so we need only two equations to
determine $a_{i},b_{i}$ only, since there is no dependence of the additive
arity $m_{i}$ exists in (\ref{an})--(\ref{ls}). Then, two consequent polyadic
powers (\ref{mn}) will give the numbers of terms $3$ and $5$. So the initial
planetext (we consider five enries $\chi=i=1,\ldots,5$) is%
\begin{equation}
T_{plain(sent)}=\left\{  t_{1},t_{2},t_{3},t_{4}t_{5}\right\}  =\left\{
a_{1},a_{2},a_{3},a_{4},a_{5}\right\}  =\left\{  11,27,17,7,28\right\}  .
\label{ta}%
\end{equation}
for the polyadic rings $\mathbb{Z}_{61,3}^{\left(  11,15\right)  }$,
$\mathbb{Z}_{85,3}^{\left(  27,28\right)  }$, $\mathbb{Z}_{181,3}^{\left(
17,18\right)  }$, $\mathbb{Z}_{73,3}^{\left(  7,8\right)  }$ and
$\mathbb{Z}_{262,3}^{\left(  28,29\right)  }$, correspondingly, while the dyad
$\tbinom{a_{i}}{m_{i}}$ plaintext becomes%
\begin{equation}
\mathbf{T}_{plain}=\left\{  \dbinom{11}{61},\dbinom{27}{85},\dbinom{17}%
{181},\dbinom{7}{73},\dbinom{28}{262}\right\}  .
\end{equation}

The polyadic amplitudes have the form (\ref{al}), and we need to find the
concrete shape of $L\left(  n_{i},\ell_{n_{i}}\right)  $ (\ref{ls}) with the
choice of the represenative (\ref{ki}). In this case the symmetric polynomials
$S_{r}$ become the standard sums of powers (Faulhaber's formula)%
\begin{equation}
S_{r}\left(  \Lambda_{i}\right)  =\frac{1}{r+1}\sum_{p=1}^{r}\dbinom{r+1}%
{p}B_{p}\Lambda_{i}^{r+1-p},\ \ \ r=1,\ldots,\Lambda_{i},
\end{equation}
where the admissable number of terms in the $n_{i}$-ary operation and
$\ell_{n_{i}}$ polyadic power (\ref{mn}) is $\Lambda_{i}=\Lambda_{i}\left(
n_{i},\ell_{n_{i}}\right)  =\ell_{n_{i}}\left(  n_{i}-1\right)  +1$,
$\Lambda_{1}=3$, $\Lambda_{2}=5$, $B_{p}$ are the Bernoulli numbers, and
$\tbinom{\bullet}{\bullet}$ are the binomial coefficients. For instance, the
first $7$ polynomials are%
\begin{align}
S_{1}\left(  \Lambda_{i}\right)   &  =\frac{1}{2}\left(  \Lambda_{i}%
^{2}+\Lambda_{i}\right)  ,\ \ \ S_{2}\left(  \Lambda_{i}\right)
=\frac{\Lambda_{i}^{3}}{3}+\frac{\Lambda_{i}^{2}}{2}+\frac{\Lambda_{i}}%
{6},\ \ \ S_{3}\left(  \Lambda_{i}\right)  =\frac{\Lambda_{i}^{4}}{4}%
+\frac{\Lambda_{i}^{3}}{2}+\frac{\Lambda_{i}^{2}}{4},\nonumber\\
S_{4}\left(  \Lambda_{i}\right)   &  =\frac{\Lambda_{i}^{5}}{5}+\frac
{\Lambda_{i}^{4}}{2}+\frac{\Lambda_{i}^{3}}{3}-\frac{\Lambda_{i}}%
{30},\ \ \ \ S_{5}\left(  \Lambda_{i}\right)  =\frac{\Lambda_{i}^{6}}{6}%
+\frac{\Lambda_{i}^{5}}{2}+\frac{5\Lambda_{i}^{4}}{12}-\frac{\Lambda_{i}^{2}%
}{12},\nonumber\\
S_{6}\left(  \Lambda_{i}\right)   &  =\frac{\Lambda_{i}^{7}}{7}+\frac
{\Lambda_{i}^{6}}{6}+\frac{\Lambda_{i}^{5}}{2}-\frac{\Lambda_{i}^{3}}{6}%
+\frac{\Lambda_{i}}{42},\ \ \ S_{7}\left(  \Lambda_{i}\right)  =\frac
{\Lambda_{i}^{8}}{8}+\frac{\Lambda_{i}^{7}}{2}+\frac{7\Lambda_{i}^{6}}%
{12}-\frac{7\Lambda_{i}^{4}}{24}+\frac{\Lambda_{i}^{2}}{12}. \label{sl}%
\end{align}

Therefore, we can write (\ref{ls}) as%
\begin{align}
L\left(  n_{i},\ell_{n_{i}}\right)   &  =a_{i}^{\Lambda_{i}-1}S_{1}\left(
\Lambda_{i}\right)  +a_{i}^{\Lambda_{i}-2}b_{i}\cdot S_{2}\left(  \Lambda
_{i}\right)  +a_{i}^{\Lambda_{i}-3}b_{i}^{2}\cdot S_{3}\left(  \Lambda
_{i}\right) \nonumber\\
&  +a_{i}^{\Lambda_{i}-4}b_{i}^{3}\cdot S_{4}\left(  \Lambda_{i}\right)
+\ldots+b_{i}^{\Lambda_{i}-1}\cdot S_{\Lambda_{i}}\left(  \Lambda_{i}\right)
, \label{ll}%
\end{align}
which fully determines the two sums of the amplitudes (\ref{al}).

It follows from (\ref{al}) and (\ref{ll}), that the multiplication amplitudes
for the polyadic powers $\ell^{\prime}=1$ and $\ell^{\prime\prime}=2$ have the
manifest form%
\begin{align}
A_{mult,\ell^{\prime}=1}^{\left(  \chi=i\right)  }  &  =a_{i}+b_{i}\left(
6a_{i}^{2}+14b_{i}a_{i}+36\right)  ,\label{am1}\\
A_{mult,\ell^{\prime}=2}^{\left(  \chi=i\right)  }  &  =a_{i}^{5}+b_{i}\left(
15a_{i}^{4}+55a_{i}^{3}b_{i}+225a_{i}^{2}b_{i}^{2}+979a_{i}b_{i}^{3}%
+4425b_{i}^{4}\right)  . \label{am2}%
\end{align}

The numerical values the amplitudes (\ref{am1})--(\ref{am2}) for the polyadic
rings $\mathbb{Z}_{61,3}^{\left(  11,15\right)  }$ ($i=1$), $\mathbb{Z}%
_{85,3}^{\left(  27,28\right)  }$ ($i=2$), $\mathbb{Z}_{181,3}^{\left(
17,18\right)  }$ ($i=3$), $\mathbb{Z}_{73,3}^{\left(  7,8\right)  }$ ($i=4$)
and $\mathbb{Z}_{262,3}^{\left(  28,29\right)  }$ ($i=5$) are%
\begin{align}
\mathbb{Z}_{61,3}^{\left(  11,15\right)  }  &  :A_{mult,\ell^{\prime}%
=1}^{\left(  \chi=i=1\right)  }=47\,411,\ \ A_{mult,\ell^{\prime}=2}^{\left(
\chi=i=1\right)  }=4017\,225\,776,\\
\mathbb{Z}_{85,3}^{\left(  27,28\right)  }  &  :A_{mult,\ell^{\prime}%
=1}^{\left(  \chi=i=2\right)  }=439515,\ \ A_{mult,\ell^{\prime}=2}^{\left(
\chi=i=2\right)  }=97\,\allowbreak090\,042\,335,\\
\mathbb{Z}_{181,3}^{\left(  17,18\right)  }  &  :A_{mult,\ell^{\prime}%
=1}^{\left(  \chi=i=3\right)  }=113\,885,\ \ A_{mult,\ell^{\prime}=2}^{\left(
\chi=i=3\right)  }=10\,\allowbreak599\,199\,955,\\
\mathbb{Z}_{73,3}^{\left(  7,8\right)  }  &  :A_{mult,\ell^{\prime}%
=1}^{\left(  \chi=i=4\right)  }=9255,\ \ A_{mult,\ell^{\prime}=2}^{\left(
\chi=i=4\right)  }=180\,225\,375,\\
\mathbb{Z}_{262,3}^{\left(  28,29\right)  }  &  :A_{mult,\ell^{\prime}%
=1}^{\left(  \chi=i=5\right)  }=489084,\ \ A_{mult,\ell^{\prime}=2}^{\left(
\chi=5\right)  }=115\,\allowbreak752\,016\,185.
\end{align}

In this way, the ciphertext (\ref{tc2}) in the dyad form sent to the recepient
by open channel is%
\begin{align}
\mathbf{T}_{cipher(mult)}  &  =\left\{  \dbinom{47\,411,4017\,225\,776}%
{61},\ \dbinom{439515,97\,\allowbreak090\,042\,335}{85},\ \right. \nonumber\\
&  \left.  \dbinom{113\,885,10\,\allowbreak599\,199\,955}{181},\ \dbinom
{9255,180\,225\,375}{73},\ \dbinom{489084,115\,\allowbreak752\,016\,185}%
{262}\right\}  .
\end{align}
The recepient knows the system of equations for parameters (\ref{am1}%
)--(\ref{am2}) for each plaintext entry $\chi=i=1,\ldots,5$, and the polyadic
powers $\ell^{\prime}=1,\ell^{\prime\prime}=2$ (\ref{mn}), then he inserts the
amplitude pairs into the system and solves it in integers for $\left(
a_{i},b_{i}\right)  $, to obtain the \textquotedblleft
hacked\textquotedblright\ ciphertext as the dyads $\tbinom{a_{i},b_{i}%
,n_{i}=3}{m_{i}}$ in the following form%
\begin{equation}
\mathbf{T}_{ckacked(mult)}=\left\{  \dbinom{11,15,3}{61},\ \dbinom
{27,28,3}{85},\ \dbinom{17,18,3}{181},\ \dbinom{7,8,3}{73},\ \dbinom
{28,29,3}{262}\right\}  .
\end{equation}

Then, the check bits $m_{i}$ (sent openly) is used by recepient to prove that
all the obtained and computed parameters actually satisfy the parameter-arity
shape mapping $\Phi\left(  a_{i},b_{i}\right)  =\left(  m_{i},n_{i}\right)  $
(\ref{f}), for smaller parameters it is in the Table 1 of
\cite{dup2017a,dup/guo}.

For consistency of the mappings $\left(  a_{i},b_{i}\right)  \mapsto\left(
m_{i},n_{i}\right)  $ we present the corresponding arity shape invariants
$I^{\left(  m_{i}\right)  }\left(  a_{i},b_{i}\right)  $, $J^{\left(
n_{i}\right)  }\left(  a_{i},b_{i}\right)  $ (\ref{i})--(\ref{j}) which should
be integer%
\begin{align}
&  \mathbb{Z}_{61,3}^{\left(  11,15\right)  }:\ \ I^{\left(  61\right)
}\left(  11,15\right)  =44,\ \ \ J^{\left(  3\right)  }\left(  11,15\right)
=88,\\
&  \mathbb{Z}_{85,3}^{\left(  27,28\right)  }:\ \ I^{\left(  85\right)
}\left(  27,28\right)  =81,\ \ \ J^{\left(  3\right)  }\left(  27,28\right)
=702,\\
&  \mathbb{Z}_{181,3}^{\left(  17,18\right)  }:\ \ I^{\left(  181\right)
}\left(  17,18\right)  =170,\ \ \ J^{\left(  3\right)  }\left(  17,18\right)
=272,\\
&  \mathbb{Z}_{73,3}^{\left(  7,8\right)  }:\ \ I^{\left(  73\right)  }\left(
7,8\right)  =63,\ \ \ J^{\left(  3\right)  }\left(  7,8\right)  =42,\\
&  \mathbb{Z}_{262,3}^{\left(  28,29\right)  }:\ \ I^{\left(  262\right)
}\left(  28,29\right)  =252,\ \ \ J^{\left(  3\right)  }\left(  28,29\right)
=756,
\end{align}

Finally, this consistency check gives the initial plaintext (\ref{ta}) as the
set of the parameters $a_{i}$ in the corresponding five polyadic rings
$\mathbb{Z}_{61,3}^{\left(  11,15\right)  }$, $\mathbb{Z}_{85,3}^{\left(
27,28\right)  }$, $\mathbb{Z}_{181,3}^{\left(  17,18\right)  }$,
$\mathbb{Z}_{73,3}^{\left(  7,8\right)  }$, $\mathbb{Z}_{262,3}^{\left(
28,29\right)  }$%
\begin{equation}
T_{plain(received)}=\left\{  11,27,17,7,28\right\}  =T_{plain(sent)}.
\end{equation}

\end{example}

The above examples show that the effectiveness of the proposed encryption and
decryption procedure is based on the complicated behavior of the
parameter-to-arity mapping $\Phi$ (\ref{f}), which is not one-to-one
(isomorphism), but it is not unique and multivalued, also it is non-injective
and non-surjective.

\section{\textbf{Conclusion}}

This article has introduced and developed a novel cryptographic framework
based on the algebraic structure of nonderived polyadic rings. By generalizing
classical binary operations to higher-arity, nonderived addition and
multiplication, we shift the foundation of encryption from well-understood
algebraic settings into a domain with profoundly more complex and less
explored structural properties.

The core of this framework is the construction of polyadic
integers---congruence classes of ordinary integers endowed with closed $m$-ary
addition and $n$-ary multiplication. The critical mathematical innovation is
the parameter-to-arity mapping $\Phi(a,b)=(m,n)$, which links the parameters
defining a congruence class to the specific arities required for algebraic
closure. This mapping is inherently intricate: it is non-injective,
non-surjective, and often multivalued. A single pair $(a,b)$ can correspond to
multiple valid arity pairs $(m,n)$, and conversely, a given arity pair can
arise from multiple distinct parameter pairs. This intricate, non-unique
relationship is not a flaw but a foundational feature, creating a vast and
labyrinthine key space.

We have detailed two concrete encryption procedures that leverage this complexity:

\begin{enumerate}
\item Encryption via summation of polyadically quantized analog signals, where
the plaintext is encoded in the additive arity $m_{i}$. 

\item Encryption via multiplication of polyadically quantized analog signals,
where the plaintext is encoded in the ring parameter $a_{i}$.
\end{enumerate}

In both methods, information is transmitted openly via signal amplitudes
derived from polyadic operations. However, successful decryption requires not
only these intercepted amplitudes but also precise knowledge of the specific
polyadic powers $(\ell_{m_{i}}, \ell_{n_{i}})$ and the functional dependencies
of representative elements used during encoding---information that is
deliberately withheld from the transmission. For a legitimate recipient with
the correct key (this auxiliary knowledge), the system reduces to a solvable
set of equations. For an adversary lacking it, the problem becomes
exceptionally difficult, as it involves solving Diophantine equations within
an unfamiliar algebraic system where the very rules of combination are
``quantized'' and non-standard.

The proposed cryptosystem offers a significant potential increase in
cryptographic security. The complexity of the $\Phi$-mapping, combined with
the flexibility in choosing polyadic powers and representative functions,
makes brute-force attacks computationally infeasible due to the explosive
growth of the key space. Furthermore, algebraic cryptanalysis is severely
hampered because the underlying nonderived polyadic structures defy the linear
and homomorphic properties commonly exploited in attacks on schemes based on
traditional binary algebras.

This work lays the essential theoretical groundwork for polyadic encryption,
demonstrating its feasibility through explicit constructions and worked
examples. It opens a promising avenue for developing robust, next-generation
cryptographic protocols that derive their strength from the rich and complex
world of higher-arity algebraic structures. Future research will focus on
implementing these schemes, analyzing their concrete security against known
attack models, and optimizing their performance for practical applications.